\documentclass{appolb}

\usepackage{graphicx}
\usepackage{hyperref}
\usepackage[sort&compress,numbers,merge]{natbib}

\begin{document}

\title{Large-$N_c$ Regge spectroscopy%
\thanks{Talk presented by WB at EEF70, Workshop on Unquenched Hadron 
Spectroscopy:
Non-Perturbative Models and Methods of QCD vs. Experiment,
Coimbra, Portugal, 1-5 September 2014}}

\author{Wojciech Broniowski
\address{Institute of Physics, Jan Kochanowski University, 25-406 Kielce, 
Poland \\ The H. Niewodnicza\'nski Institute of Nuclear Physics, Polish
Academy of Sciences, PL-31342 Cracow, Poland \\~}
\\
Enrique Ruiz Arriola
\address{Departamento de F\'{\i}sica At\'{o}mica, Molecular y Nuclear and 
Instituto Carlos I de  F{\'\i}sica Te\'orica y Computacional, 
 Universidad de Granada, E-18071 Granada, Spain \\~}
\\
Pere Masjuan
\address{PRISMA Cluster of Excellence, Institut f\"ur Kernphysik, Johannes 
Gutenberg-Universt\"at, Mainz D-55099, Germany}}

\maketitle

\begin{abstract}
This talk, dedicated to Eef van Beveren on the occasion of his birthday,
reviews some of our results concerning the hadron
spectroscopy, Regge trajectories, and the large-$N_c$ meson-dominance of 
hadronic form factors.
\end{abstract}


\bigskip

\begin{figure}[tb]
\centerline{%
\includegraphics[width=0.47\textwidth]{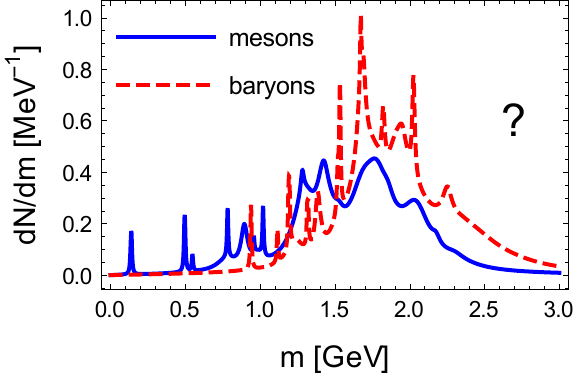} \hfill
\includegraphics[width=0.475\textwidth]{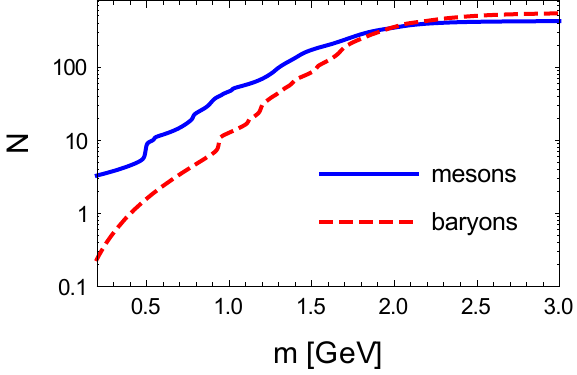}
}
\caption{Left: excitation function of QCD for light (u,d,s) hadrons, where the 
states are represented as 
Bright-Wigner distributions. Right: the corresponding number of states below 
mass $m$. The question mark indicates the experimentally unexplored region.
\label{fig:excit}}
\end{figure}

\begin{figure}[tb]
\centerline{%
\includegraphics[width=0.5\textwidth]{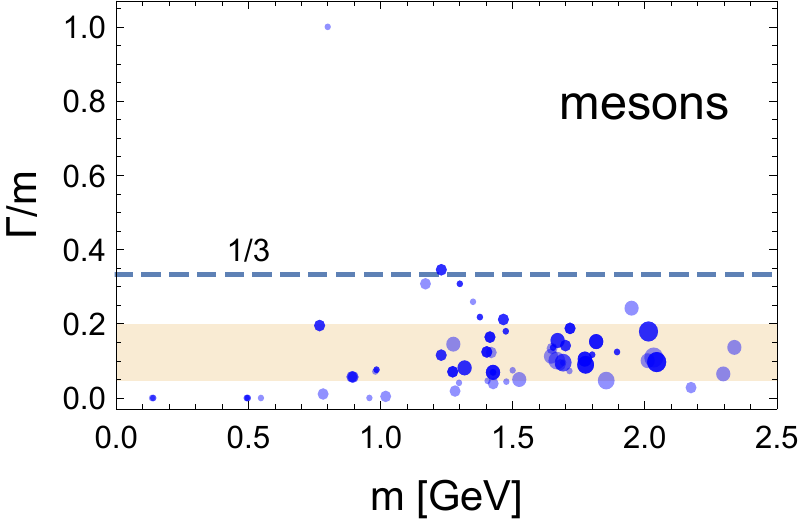}%
\includegraphics[width=0.5\textwidth]{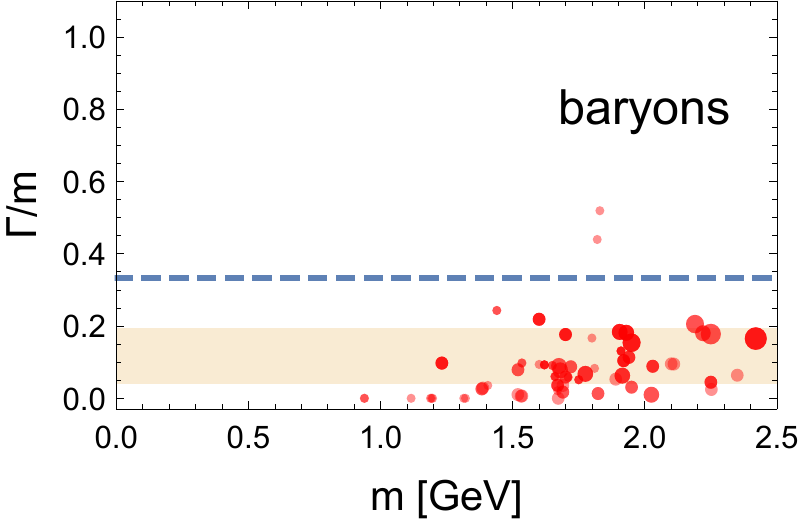}
}
\caption{Ratio of the width to mass for the known (u,d,s) meson (left) and 
baryon (right) states. The size of the dots is proportional to the spin 
degeneracy, and the intensity to the isospin degeneracy. The band indicates the 
mean $\pm$ standard deviation. 
\label{fig:wom}}
\end{figure}

The presentation is based on several of our recent 
papers~\cite{RuizArriola:2010fj,Arriola:2010aj,Masjuan:2012gc,%
Masjuan:2013xta,Masjuan:2012sk} devoted to hadron
spectroscopy involving the large-$N_c$ arguments and Regge trajectories.
We begin with a brief review of the old Hagedorn~\cite{Hagedorn:1965st} 
idea, applied to the fundamental question of understanding the spectrum of QCD, 
as well as its applications to thermodynamic properties which find practical 
use in understanding the lattice data and modeling the relativistic heavy-ion 
collisions.
The excitation function of QCD is presented in Fig.~\ref{fig:excit}, where we 
plot the density of states (left panel) represented via the Breit-Wigner 
functions (for plotting purposes the stable states were attributed some finite 
width), as well as the cumulative number of states with mass below $m$. We 
include all light hadron states as described 
in~\cite{Broniowski:2000bj,Broniowski:2004yh}. We note that the cumulative 
number of states increases exponentially up to $m \sim 2$~GeV, in accordance to 
the Hagedorn hypothesis. Above, no states have been identified, which may be a 
feature of QCD (the states are wider as $m$ grows), or result from no 
experimental ``coverage''. The mentioned growth of the width of resonances with 
the mass is visualized in Fig.~\ref{fig:wom}. Except for some unusual states, 
such as the $\sigma$ meson, all states follow this trend. Thus, as $m$ 
increases, we have more and more states which become wider. This leads to the 
expectation that the region indicated with a question mark in left 
Fig.~\ref{fig:excit} is indeed filled with strength coming from such states.

\begin{figure}[tb]
\begin{center}
\includegraphics[width=.52\textwidth]{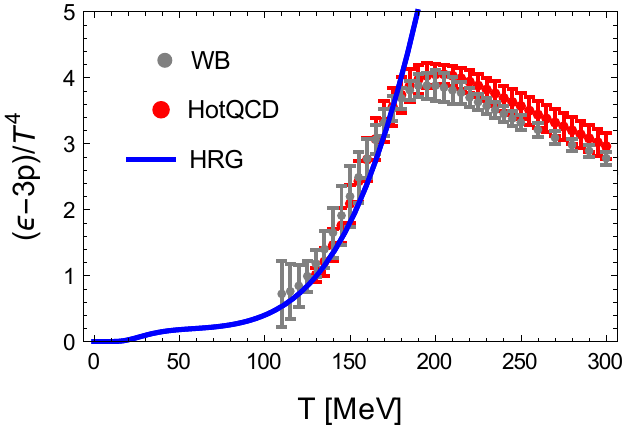}%
\hspace{.03\textwidth}%
\begin{minipage}[b]{.45\textwidth}\caption{ \small\label{fig:trAn} 
\small 
Trace anomaly of QCD from the hadron resonance gas (HRG) compared to the 
lattice data of the Wuppertal-Budapest (WB)~\cite{Borsanyi:2013bia} and 
HotQCD~\cite{Bazavov:2014pvz} collaborations.
}
\end{minipage}
\end{center}
\end{figure}

On may use the resonances to evaluate thermodynamic properties of QCD in 
the hadronic phase (see, e.g., \cite{Arriola:2014bfa} and references therein). 
An example of the trace anomaly $\Theta^\mu_\mu$ is shown in 
Fig.~\ref{fig:trAn}, where we can enjoy the nice display of the parton-hadron 
duality in the agreement of the hadron resonance gas and the lattice data. 

As discussed in~\cite{Broniowski:2000hd} (see also references therein), the 
Hagedorn growth of states may be explained with string models which result in 
large degeneracy of the daughter Regge trajectories. These models also explain 
the faster growth for baryons than for mesons, as realized in the data. Thus 
the question of the Regge spectra is immanently related to the issues discussed 
here. 

\begin{figure}[tb]
\begin{center}
\includegraphics[width=.52\textwidth]{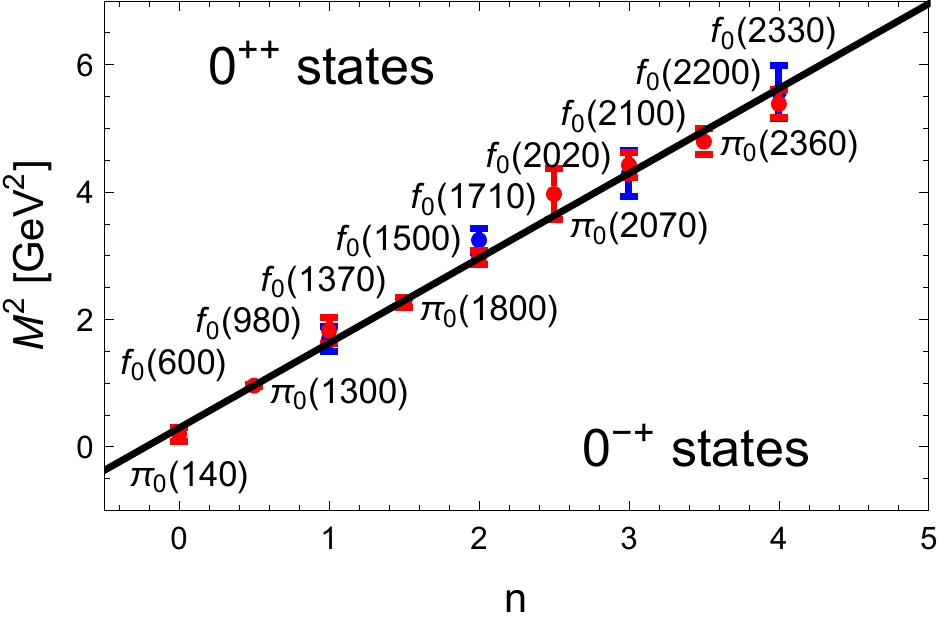}%
\hspace{.03\textwidth}%
\begin{minipage}[b]{.45\textwidth}\caption{ \small\label{fig:sp} 
\small 
Radial Regge trajectories for the $f_0$ and pion states. The two $f_0$ 
trajectories were overlaid, yielding a single trajectory with half the slope. 
The universality with the pion trajectory is vivid.
}
\end{minipage}
\end{center}
\end{figure}

A comprehensive review of the Regge spectroscopy in both the angular and radial 
quantum numbers is presented in~\cite{Masjuan:2012gc}, 
following the compilation of~\cite{Anisovich:2000kxa}. An example of special 
interest to Eef should be the case of scalar-isoscalar states. It turns out 
that these states may be arranged along two Regge trajectories with the 
standard slope, or, equivalently, on one with half the slope. Then it becomes 
degenerate with the pion trajectory, as seen from Fig.~\ref{fig:sp}, 
cf.~\cite{RuizArriola:2010fj}.

\begin{figure}[tb]
\centerline{%
\includegraphics[width=0.5\textwidth]{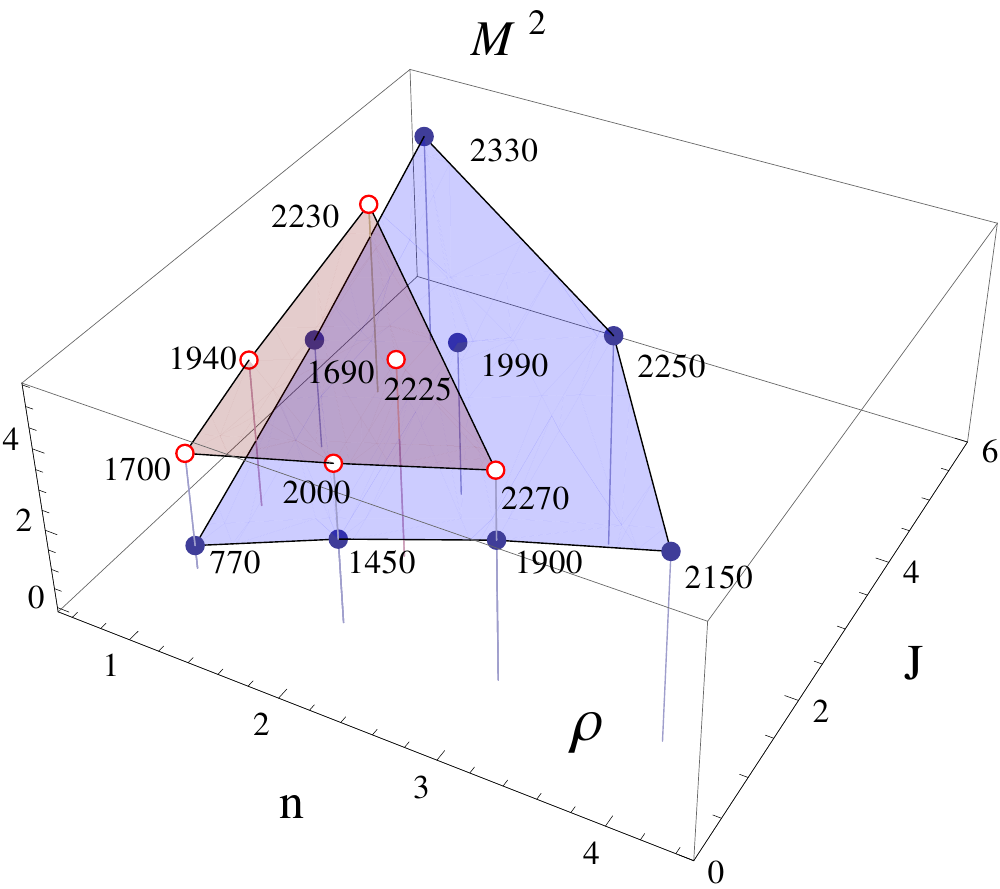}%
\includegraphics[width=0.5\textwidth]{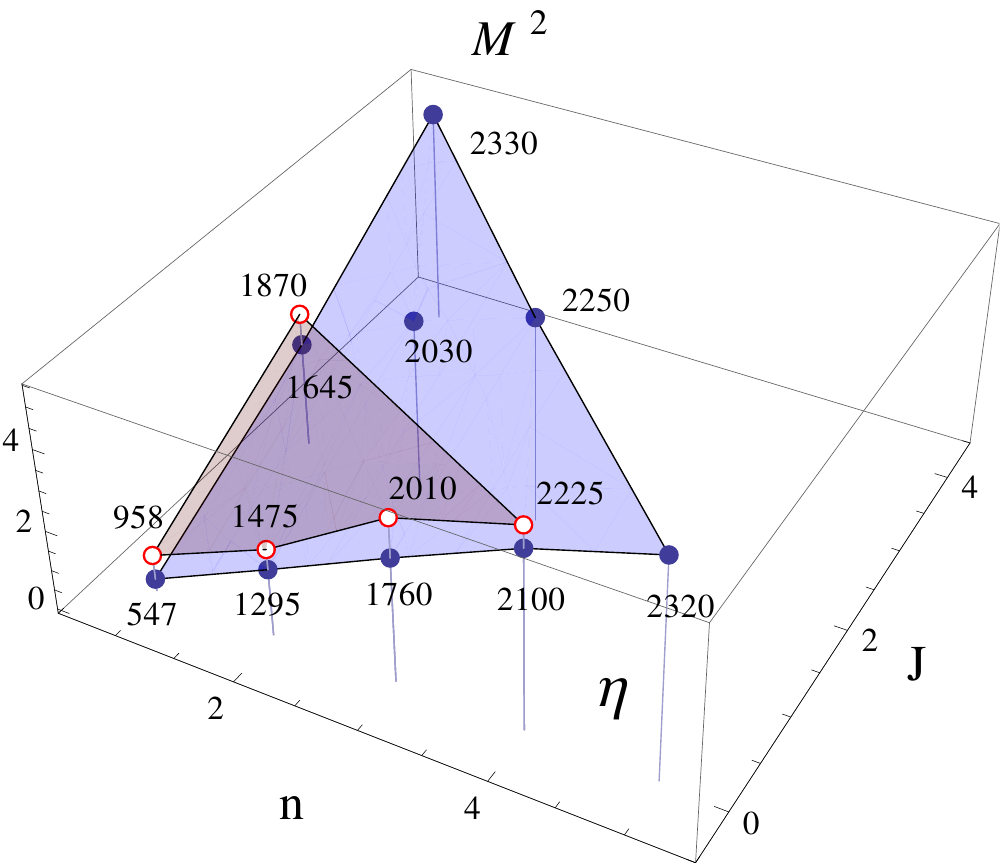}
}
\caption{Regge planes based on the angular and radial $\rho$ and $\eta$ Regge 
trajectories.
\label{fig:plane}}
\end{figure}

\begin{figure}[tb]
\centerline{%
\includegraphics[width=0.65\textwidth]{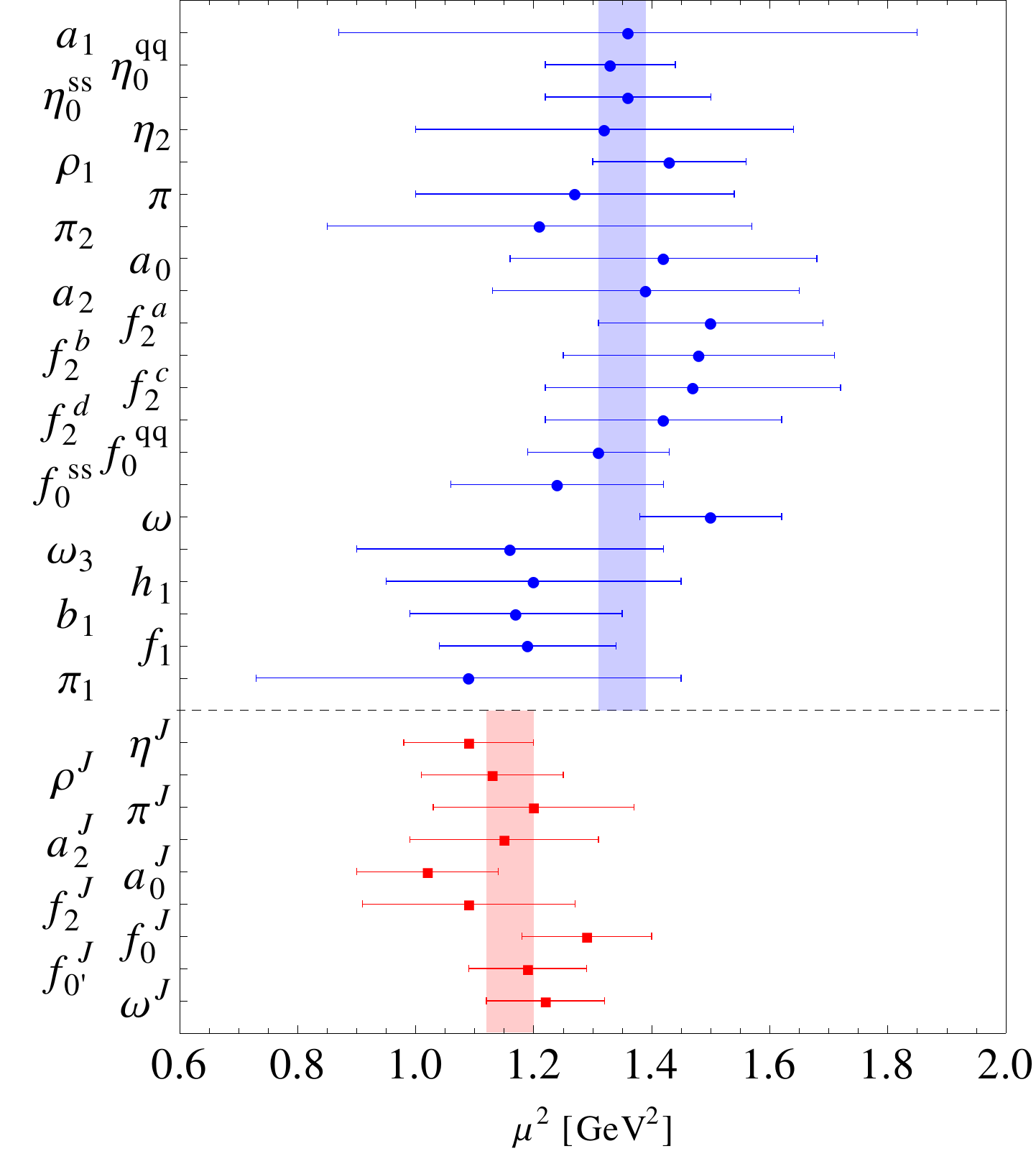}
}
\caption{Slope parameters $\mu^2=a$ for the radial and $\mu^2=b$ for the 
angular Regge trajectories from Ref.~\cite{Masjuan:2012gc}. The bands indicate 
the mean $\pm$ standard deviation. The lack of universality of the radial and 
angular slopes is evident. 
\label{fig:slope}}
\end{figure}

As hadrons may be grouped in angular and radial Regge trajectories, both may be 
combined into {\em Regge planes}, where the principal and daughter planes are 
nearly parallel. An example of two such cases is shown in Fig.~\ref{fig:plane}.
A fundamental question here is whether the slopes of the radial and angular 
Regge trajectories are universal or not. Our study presented in 
Ref.~\cite{Masjuan:2012gc} shows that at the statistical level of 4.5 
standard deviations there is no universality. In this study we have fitted 
the formula 
\begin{eqnarray}
M^2=a n + b J + c = 1.38(4)n+1.12(4)J-1.25(4). \label{eq:regge}
\end{eqnarray}
The result of the fit for the individual trajectories is presented in 
Fig.~\ref{fig:slope}.

An interesting application of meson spectroscopy is the explanation of various 
hadronic form factors with the meson dominance principle, which generalizes 
the well-known vector-meson dominance. In the large-$N_c$ limit, where all 
diagrams are trees, one proceeds by coupling the mesons with appropriate 
quantum numbers to currents, and then to the probed hadron. The masses of 
the mesons are taken as their physical values, while the relevant coupling 
constants are fitted to the form factor data. The QCD short-distance 
constraints may be satisfied when a sufficient number of mesons is 
included~\cite{Pich:2002xy,Masjuan:2012sk}. Of course, one may always include 
more states, even infinitely many.

\begin{figure}[htb]
\centerline{%
\includegraphics[width=0.5\textwidth]{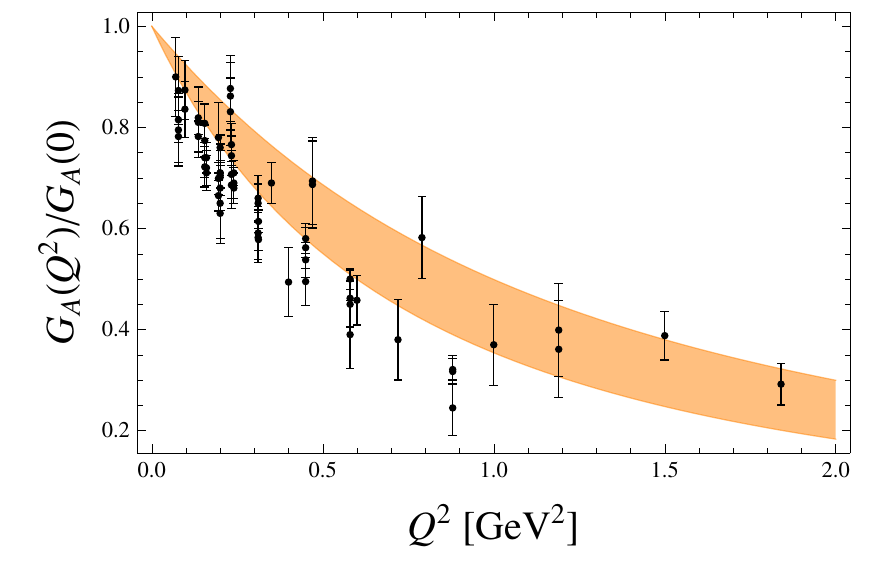}%
\includegraphics[width=0.5\textwidth]{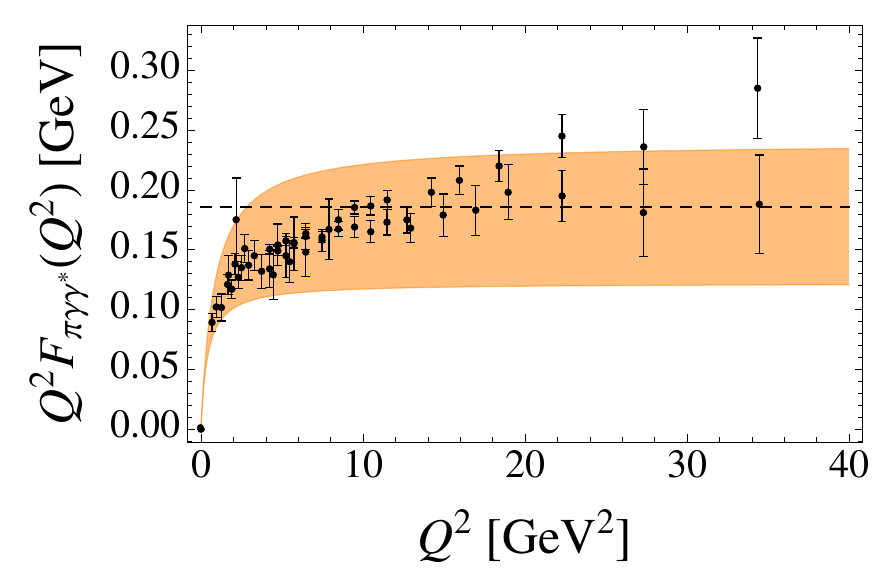}
}
\caption{Examples of the meson-dominance principle: (left) the axial-vector 
form factor of the nucleon compared to the lattice 
data~\cite{Alexandrou:2010hf} and (right) the pion-photon transition form 
factor multiplied by $Q^2$, compared to the experimental 
data~\cite{Behrend:1990sr,Gronberg:1997fj,Aubert:2009mc,Uehara:2012ag}. The 
model uncertainty bands are obtained with the $1/N_c$ rule.
\label{fig:tr}}
\end{figure}

An important issue in such 
studies is the error analysis, which allows for an assessment how well the 
approach works. In our studies, based on the large-$N_c$ limit, we have used 
either the width of the state as its uncertainty of mass, or mass divided by 
$N_c$. We call them, correspondingly, the $\Gamma$ rule or the $1/N_c$ rule.
In Fig.~\ref{fig:tr} we show two examples of such calculations: the 
axial-vector isovector form factor of the nucleon and the pion-photon 
transition form factor. The bands are obtained from the $1/N_c$ rule. We note 
that the data are compatible with the experimental data. A similar agreement is 
found for other form factors of the nucleon and the pion (electromagnetic, 
gravitational)~\cite{Masjuan:2012sk}, indicating that the large-$N_c$ 
meson-dominance principle is capable of reproducing the data.

\bigskip

Happy birthday, Eef!

\bigskip

The authors acknowledge the support of the Polish National Science 
Centre, grants DEC-2011/01/B/ST2/03915 and DEC-2012/06/A/ST2/00390 (WB), 
Spanish Mineco, grant No. FIS2011-24149, and Junta de Andalucía, 
grant No.~FQM225 (ERA).

\small

\bibliography{fromhep}

\begin{thebibliography}{10}%
\makeatletter
\providecommand \@ifxundefined [1]{%
 \ifx #1\undefined \expandafter \@firstoftwo
 \else \expandafter \@secondoftwo
\fi
}%
\providecommand \@ifnum [1]{%
 \ifnum #1\expandafter \@firstoftwo
 \else \expandafter \@secondoftwo
\fi
}%
\providecommand \enquote [1]{``#1''}%
\providecommand \bibnamefont  [1]{#1}%
\providecommand \bibfnamefont [1]{#1}%
\providecommand \citenamefont [1]{#1}%
\providecommand\href[0]{\@sanitize\@href}%
\providecommand\@href[1]{\endgroup\@@startlink{#1}\endgroup\@@href}%
\providecommand\@@href[1]{#1\@@endlink}%
\providecommand \@sanitize [0]{\begingroup\catcode`\&12\catcode`\#12\relax}%
\@ifxundefined \pdfoutput {\@firstoftwo}{%
 \@ifnum{\z@=\pdfoutput}{\@firstoftwo}{\@secondoftwo}%
}{%
 \providecommand\@@startlink[1]{\leavevmode\special{html:<a href="#1">}}%
 \providecommand\@@endlink[0]{\special{html:</a>}}%
}{%
 \providecommand\@@startlink[1]{%
  \leavevmode
  \pdfstartlink
   attr{/Border[0 0 1 ]/H/I/C[0 1 1]}%
   user{/Subtype/Link/A<</Type/Action/S/URI/URI(#1)>>}%
  \relax
 }%
 \providecommand\@@endlink[0]{\pdfendlink}%
}%
\providecommand \url  [0]{\begingroup\@sanitize \@url }%
\providecommand \@url [1]{\endgroup\@href {#1}{\urlprefix}}%
\providecommand \urlprefix [0]{URL }%
\providecommand \Eprint[0]{\href }%
\@ifxundefined \urlstyle {%
  \providecommand \doi [1]{doi:\discretionary{}{}{}#1}%
}{%
  \providecommand \doi [0]{doi:\discretionary{}{}{}\begingroup
  \urlstyle{rm}\Url }%
}%
\providecommand \doibase [0]{http://dx.doi.org/}%
\providecommand \Doi[1]{\href{\doibase#1}}%
\providecommand \bibAnnote [3]{%
  \BibitemShut{#1}%
  \begin{quotation}\noindent
    \textsc{Key:}\ #2\\\textsc{Annotation:}\ #3%
  \end{quotation}%
}%
\providecommand \bibAnnoteFile [2]{%
  \IfFileExists{#2}{\bibAnnote {#1} {#2} {\input{#2}}}{}%
}%
\providecommand \typeout [0]{\immediate \write \m@ne }%
\providecommand \selectlanguage [0]{\@gobble}%
\providecommand \bibinfo [0]{\@secondoftwo}%
\providecommand \bibfield [0]{\@secondoftwo}%
\providecommand \translation [1]{[#1]}%
\providecommand \BibitemOpen[0]{}%
\providecommand \bibitemStop [0]{}%
\providecommand \bibitemNoStop [0]{.\EOS\space}%
\providecommand \EOS [0]{\spacefactor3000\relax}%
\providecommand \BibitemShut [1]{\csname bibitem#1\endcsname}%
\bibitem{RuizArriola:2010fj}%
  \BibitemOpen
  \bibfield{author}{%
  \bibinfo {author} {\bibfnamefont{E.}~\bibnamefont{Ruiz~Arriola}}\ and\
  \bibinfo {author} {\bibfnamefont{W.}~\bibnamefont{Broniowski}},\ }%
  \bibfield{journal}{%
  \Doi{10.1103/PhysRevD.81.054009}{\bibinfo {journal} {Phys.Rev.}}\ }%
  \textbf{\bibinfo {volume} {D81}},\ \bibinfo {pages} {054009} (\bibinfo {year}
  {2010})%
  \bibAnnoteFile{NoStop}{RuizArriola:2010fj}%
\bibitem{Arriola:2010aj}%
  \BibitemOpen
  \bibfield{author}{%
  \bibinfo {author} {\bibfnamefont{E.}~\bibnamefont{Ruiz~Arriola}}\ and\
  \bibinfo {author} {\bibfnamefont{W.}~\bibnamefont{Broniowski}},\ }%
  \bibfield{journal}{%
  \Doi{10.1063/1.3575030}{\bibinfo {journal} {AIP Conf.Proc.}}\ }%
  \textbf{\bibinfo {volume} {1343}},\ \bibinfo {pages} {361} (\bibinfo {year}
  {2011})%
  \bibAnnoteFile{NoStop}{Arriola:2010aj}%
\bibitem{Masjuan:2012gc}%
  \BibitemOpen
  \bibfield{author}{%
  \bibinfo {author} {\bibfnamefont{P.}~\bibnamefont{Masjuan}}, \bibinfo
  {author} {\bibfnamefont{E.}~\bibnamefont{Ruiz~Arriola}},\ and\ \bibinfo
  {author} {\bibfnamefont{W.}~\bibnamefont{Broniowski}},\ }%
  \bibfield{journal}{%
  \Doi{10.1103/PhysRevD.85.094006}{\bibinfo {journal} {Phys.Rev.}}\ }%
  \textbf{\bibinfo {volume} {D85}},\ \bibinfo {pages} {094006} (\bibinfo {year}
  {2012})%
  \bibAnnoteFile{NoStop}{Masjuan:2012gc}%
\bibitem{Masjuan:2013xta}%
  \BibitemOpen
  \bibfield{author}{%
  \bibinfo {author} {\bibfnamefont{P.}~\bibnamefont{Masjuan}}, \bibinfo
  {author} {\bibfnamefont{E.}~\bibnamefont{Ruiz~Arriola}},\ and\ \bibinfo
  {author} {\bibfnamefont{W.}~\bibnamefont{Broniowski}},\ }%
  \bibfield{journal}{%
  \Doi{10.1103/PhysRevD.87.118502}{\bibinfo {journal} {Phys.Rev.}}\ }%
  \textbf{\bibinfo {volume} {D87}},\ \bibinfo {pages} {118502} (\bibinfo {year}
  {2013})%
  \bibAnnoteFile{NoStop}{Masjuan:2013xta}%
\bibitem{Masjuan:2012sk}%
  \BibitemOpen
  \bibfield{author}{%
  \bibinfo {author} {\bibfnamefont{P.}~\bibnamefont{Masjuan}}, \bibinfo
  {author} {\bibfnamefont{E.}~\bibnamefont{Ruiz~Arriola}},\ and\ \bibinfo
  {author} {\bibfnamefont{W.}~\bibnamefont{Broniowski}},\ }%
  \bibfield{journal}{%
  \Doi{10.1103/PhysRevD.87.014005}{\bibinfo {journal} {Phys.Rev.}}\ }%
  \textbf{\bibinfo {volume} {D87}},\ \bibinfo {pages} {014005} (\bibinfo {year}
  {2013})%
  \bibAnnoteFile{NoStop}{Masjuan:2012sk}%
\bibitem{Hagedorn:1965st}%
  \BibitemOpen
  \bibfield{author}{%
  \bibinfo {author} {\bibfnamefont{R.}~\bibnamefont{Hagedorn}},\ }%
  \bibfield{journal}{%
  \bibinfo {journal} {Nuovo Cim.Suppl.}\ }%
  \textbf{\bibinfo {volume} {3}},\ \bibinfo {pages} {147} (\bibinfo {year}
  {1965})%
  \bibAnnoteFile{NoStop}{Hagedorn:1965st}%
\bibitem{Broniowski:2000bj}%
  \BibitemOpen
  \bibfield{author}{%
  \bibinfo {author} {\bibfnamefont{W.}~\bibnamefont{Broniowski}}\ and\ \bibinfo
  {author} {\bibfnamefont{W.}~\bibnamefont{Florkowski}},\ }%
  \bibfield{journal}{%
  \Doi{10.1016/S0370-2693(00)00992-8}{\bibinfo {journal} {Phys.Lett.}}\ }%
  \textbf{\bibinfo {volume} {B490}},\ \bibinfo {pages} {223} (\bibinfo {year}
  {2000})%
  \bibAnnoteFile{NoStop}{Broniowski:2000bj}%
\bibitem{Broniowski:2004yh}%
  \BibitemOpen
  \bibfield{author}{%
  \bibinfo {author} {\bibfnamefont{W.}~\bibnamefont{Broniowski}}, \bibinfo
  {author} {\bibfnamefont{W.}~\bibnamefont{Florkowski}},\ and\ \bibinfo
  {author} {\bibfnamefont{L.~Y.}\ \bibnamefont{Glozman}},\ }%
  \bibfield{journal}{%
  \Doi{10.1103/PhysRevD.70.117503}{\bibinfo {journal} {Phys.Rev.}}\ }%
  \textbf{\bibinfo {volume} {D70}},\ \bibinfo {pages} {117503} (\bibinfo {year}
  {2004})%
  \bibAnnoteFile{NoStop}{Broniowski:2004yh}%
\bibitem{Arriola:2014bfa}%
  \BibitemOpen
  \bibfield{author}{%
  \bibinfo {author} {\bibfnamefont{E.~R.}\ \bibnamefont{Arriola}}, \bibinfo
  {author} {\bibfnamefont{L.}~\bibnamefont{Salcedo}},\ and\ \bibinfo {author}
  {\bibfnamefont{E.}~\bibnamefont{Megias}}}%
   (\bibinfo {year} {2014}),\
  \Eprint{http://arxiv.org/abs/1410.3869}{arXiv:1410.3869 [hep-ph]}%
  \bibAnnoteFile{NoStop}{Arriola:2014bfa}%
\bibitem{Borsanyi:2013bia}%
  \BibitemOpen
  \bibfield{author}{%
  \bibinfo {author} {\bibfnamefont{S.}~\bibnamefont{Borsanyi}}, \bibinfo
  {author} {\bibfnamefont{Z.}~\bibnamefont{Fodor}}, \bibinfo {author}
  {\bibfnamefont{C.}~\bibnamefont{Hoelbling}}, \bibinfo {author}
  {\bibfnamefont{S.~D.}\ \bibnamefont{Katz}}, \bibinfo {author}
  {\bibfnamefont{S.}~\bibnamefont{Krieg}}, \emph{et~al.},\ }%
  \bibfield{journal}{%
  \Doi{10.1016/j.physletb.2014.01.007}{\bibinfo {journal} {Phys.Lett.}}\ }%
  \textbf{\bibinfo {volume} {B730}},\ \bibinfo {pages} {99} (\bibinfo {year}
  {2014})%
  \bibAnnoteFile{NoStop}{Borsanyi:2013bia}%
\bibitem{Bazavov:2014pvz}%
  \BibitemOpen
  \bibfield{author}{%
  \bibinfo {author} {\bibfnamefont{A.}~\bibnamefont{Bazavov}} \emph{et~al.}
  (\bibinfo {collaboration} {HotQCD Collaboration}),\ }%
  \bibfield{journal}{%
  \Doi{10.1103/PhysRevD.90.094503}{\bibinfo {journal} {Phys.Rev.}}\ }%
  \textbf{\bibinfo {volume} {D90}},\ \bibinfo {pages} {094503} (\bibinfo {year}
  {2014})%
  \bibAnnoteFile{NoStop}{Bazavov:2014pvz}%
\bibitem{Broniowski:2000hd}%
  \BibitemOpen
  \bibfield{author}{%
  \bibinfo {author} {\bibfnamefont{W.}~\bibnamefont{Broniowski}},\ \bibinfo
  {pages} {3}}%
   (\bibinfo {year} {2000}),\
  \Eprint{http://arxiv.org/abs/hep-ph/0008112}{arXiv:hep-ph/0008112 [hep-ph]}%
  \bibAnnoteFile{NoStop}{Broniowski:2000hd}%
\bibitem{Anisovich:2000kxa}%
  \BibitemOpen
  \bibfield{author}{%
  \bibinfo {author} {\bibfnamefont{A.}~\bibnamefont{Anisovich}}, \bibinfo
  {author} {\bibfnamefont{V.}~\bibnamefont{Anisovich}},\ and\ \bibinfo {author}
  {\bibfnamefont{A.}~\bibnamefont{Sarantsev}},\ }%
  \bibfield{journal}{%
  \Doi{10.1103/PhysRevD.62.051502}{\bibinfo {journal} {Phys.Rev.}}\ }%
  \textbf{\bibinfo {volume} {D62}},\ \bibinfo {pages} {051502} (\bibinfo {year}
  {2000})%
  \bibAnnoteFile{NoStop}{Anisovich:2000kxa}%
\bibitem{Pich:2002xy}%
  \BibitemOpen
  \bibfield{author}{%
  \bibinfo {author} {\bibfnamefont{A.}~\bibnamefont{Pich}},\ \bibinfo {pages}
  {239}}%
   (\bibinfo {year} {2002}),\
  \Eprint{http://arxiv.org/abs/hep-ph/0205030}{arXiv:hep-ph/0205030 [hep-ph]}%
  \bibAnnoteFile{NoStop}{Pich:2002xy}%
\bibitem{Alexandrou:2010hf}%
  \BibitemOpen
  \bibfield{author}{%
  \bibinfo {author} {\bibfnamefont{C.}~\bibnamefont{Alexandrou}} \emph{et~al.}
  (\bibinfo {collaboration} {ETM Collaboration}),\ }%
  \bibfield{journal}{%
  \Doi{10.1103/PhysRevD.83.045010}{\bibinfo {journal} {Phys.Rev.}}\ }%
  \textbf{\bibinfo {volume} {D83}},\ \bibinfo {pages} {045010} (\bibinfo {year}
  {2011})%
  \bibAnnoteFile{NoStop}{Alexandrou:2010hf}%
\bibitem{Behrend:1990sr}%
  \BibitemOpen
  \bibfield{author}{%
  \bibinfo {author} {\bibfnamefont{H.}~\bibnamefont{Behrend}} \emph{et~al.}
  (\bibinfo {collaboration} {CELLO Collaboration}),\ }%
  \bibfield{journal}{%
  \Doi{10.1007/BF01549692}{\bibinfo {journal} {Z.Phys.}}\ }%
  \textbf{\bibinfo {volume} {C49}},\ \bibinfo {pages} {401} (\bibinfo {year}
  {1991})%
  \bibAnnoteFile{NoStop}{Behrend:1990sr}%
\bibitem{Gronberg:1997fj}%
  \BibitemOpen
  \bibfield{author}{%
  \bibinfo {author} {\bibfnamefont{J.}~\bibnamefont{Gronberg}} \emph{et~al.}
  (\bibinfo {collaboration} {CLEO Collaboration}),\ }%
  \bibfield{journal}{%
  \Doi{10.1103/PhysRevD.57.33}{\bibinfo {journal} {Phys.Rev.}}\ }%
  \textbf{\bibinfo {volume} {D57}},\ \bibinfo {pages} {33} (\bibinfo {year}
  {1998})%
  \bibAnnoteFile{NoStop}{Gronberg:1997fj}%
\bibitem{Aubert:2009mc}%
  \BibitemOpen
  \bibfield{author}{%
  \bibinfo {author} {\bibfnamefont{B.}~\bibnamefont{Aubert}} \emph{et~al.}
  (\bibinfo {collaboration} {BaBar Collaboration}),\ }%
  \bibfield{journal}{%
  \Doi{10.1103/PhysRevD.80.052002}{\bibinfo {journal} {Phys.Rev.}}\ }%
  \textbf{\bibinfo {volume} {D80}},\ \bibinfo {pages} {052002} (\bibinfo {year}
  {2009})%
  \bibAnnoteFile{NoStop}{Aubert:2009mc}%
\bibitem{Uehara:2012ag}%
  \BibitemOpen
  \bibfield{author}{%
  \bibinfo {author} {\bibfnamefont{S.}~\bibnamefont{Uehara}} \emph{et~al.}
  (\bibinfo {collaboration} {Belle Collaboration}),\ }%
  \bibfield{journal}{%
  \Doi{10.1103/PhysRevD.86.092007}{\bibinfo {journal} {Phys.Rev.}}\ }%
  \textbf{\bibinfo {volume} {D86}},\ \bibinfo {pages} {092007} (\bibinfo {year}
  {2012})%
  \bibAnnoteFile{NoStop}{Uehara:2012ag}%
\end{thebibliography}%

\end{document}